\documentstyle[12pt,epsfig]{article}
\begin{document}
%
\def    \be             {\begin{equation}}
\def    \ee             {\end{equation}}
\def\eq#1{{eq. (\ref{#1})}}
\newcommand{\ptmis}{{ {\rm p} \hspace{-0.53 em} \raisebox{-0.27 ex}{/}_T }}
%

\font\fortssbx=cmssbx10 scaled \magstep2
\hbox to \hsize{
\hbox{\fortssbx University of Wisconsin - Madison}
\hfill$\vcenter{\hbox{\bf MADPH-95-923}
                \hbox{\bf FTUV/95-69}
                \hbox{\bf IFIC/95-72}
                \hbox{\bf KA-TP-11-1995}
                \hbox{\bf hep-ph/9512362}
                \hbox{December 1995}}$ }

\vskip 1.25cm

\begin{center}
{\bf LOOKING FOR INVISIBLY DECAYING HIGGS BOSONS
THROUGH THE FINAL STATE $b\bar{b}+\protect\ptmis$}
\footnote{Contribution to the workshop on Physics at LEP2, Higgs Physics
Group, M.\ Carena and P.\ Zerwas conveners.}\\

\vskip 0.5cm

{\bf O.\ J.\ P.\ \'Eboli$^{a,1}$, F.\ de Campos$^{b}$, J.\ Rosiek$^{c}$
and J.\ W.\ F.\ Valle$^{b}$}\\

\vskip 0.4cm

$^a$Physics Department, University of Wisconsin, Madison, WI 53706, USA\\

\vskip 0.2cm

$^b$Instituto de F\'{\i}sica Corpuscular - C.S.I.C., Dept.\ de F\'{\i}sica
Te\`orica,\\
 Universitat de Val\`encia 46100 Burjassot, Val\`encia, Spain\\

\vskip 0.2cm

$^c$Institut f\"ur Theoretische Physik, Universit\"at Karlsruhe\\
Postfach 6980, 76128 Karlsruhe, Germany\\

\vskip 1cm

\begin{quote}
\protect\small \baselineskip 12pt
We study the potential of LEP II to unravel the existence of invisibly
decaying Higgs bosons through the reaction $e^+ e^- \rightarrow b
\bar{b}+ \ptmis$. We perform our analyses in a model independent way
and our results show that LEP II is capable of discovering such a
Higgs for a wide range of masses and couplings.
\end{quote}
\end{center}

There are a variety of well motivated extensions of the standard
model (SM) with an spontaneously broken
global symmetry. This symmetry could be either be lepton number
or a combination of family lepton numbers \cite{CMP,majoron}.
These models are characterised by a more complex symmetry
breaking sector which contain additional Higgs bosons.
It is specially interesting for our purposes to consider models
where such symmetry is broken at the electroweak scale \cite{HJJ,MASI_pot3}.
In general, these models contain a massless Goldstone boson, called majoron
($J$), which interacts very weakly with normal matter.  In such models,
the normal doublet Higgs is expected to have sizeable invisible decay
modes to the majoron, due to the strong Higgs majoron coupling.
This can have a significant effect
on the Higgs phenomenology at LEP II.  In particular, the invisible
decay could contribute to the signal of two acoplanar jets and missing
momentum. This feature of majoron models allows one to strongly
constrain the Higgs mass in spite of the occurrence of extra
parameters compared to the SM.  In particular, the LEP I limit on the
predominantly doublet Higgs mass is close to the SM limit irrespective
of the decay mode of the Higgs boson \cite{valle:1,dproy}.

In this work we consider a model containing two Higgs doublets
($\phi_{1,2}$) and a singlet ($\sigma$) under the $SU(2)_L \times
U(1)_Y$ group.  The singlet Higgs field carries a non-vanishing
$U(1)_L$ charge, which could be lepton number. Here we
only need to specify the scalar potential of the model:
\begin{eqnarray}
\label{N2}
V &=&\mu_{i}^2\phi^{\dagger}_i\phi_i+\mu_{\sigma}^2\sigma^{\dagger}
\sigma + \lambda_{i}(\phi^{\dagger}_i\phi_i)^2+
   \lambda_{3} (\sigma^{\dagger}\sigma)^2+
\nonumber\\
& &\lambda_{12}(\phi^{\dagger}_1\phi_1)(\phi^{\dagger}_2\phi_2)
+\lambda_{13}(\phi^{\dagger}_1\phi_1)(\sigma^{\dagger}\sigma)+
\lambda_{23}(\phi^{\dagger}_2\phi_2)(\sigma^{\dagger}\sigma)
\nonumber\\
& & +\delta(\phi^{\dagger}_1\phi_2)(\phi^{\dagger}_2\phi_1)+
    \frac{1}{2}\kappa[(\phi^{\dagger}_1\phi_2)^2+\;\;h.\;c.]
\end{eqnarray}
where the sum over repeated indices $i$=1,2 is assumed.

Minimisation of the above potential leads to the spontaneous $SU(2)_L
\times U(1)_Y \times U(1)_L$ symmetry breaking and allows us to identify
a total of three massive CP even scalars $H_{i} \:$ (i=1,2,3), plus a
massive pseudoscalar $A$ and the massless majoron $J$. We assume that
at the LEP II energies only three Higgs particles can be produced: the
lightest CP-even scalar $h$, the CP-odd massive scalar $A$, and the
massless majoron $J$. Notwithstanding, our analyses is also valid for
the situation where the Higgs boson $A$ is absent \cite{valle:2},
which can be obtained
by setting the couplings of this field to zero.

At LEP II, the main production mechanisms of invisible Higgs bosons
are the Bjorken process ($e^+e^- \rightarrow h Z$) and the associated
production of Higgs bosons pairs ($e^+e^- \rightarrow A h$), which
rely upon the couplings $hZZ$ and $hAZ$ respectively. The important
feature of the above model is that, because of its singlet nature,
the majoron is not size-ably coupled to the gauge bosons and cannot
be produced directly, therefore, thereby evading strong LEP I
constraints. The $hZZ$ and $hAZ$ couplings depend on the
model parameters via the appropriate mixing angles, but they can be
effectively expressed in terms of the two parameters $\epsilon_A$,
$\epsilon_B$:
\begin{eqnarray}
\label{HZZ3}
{\cal L}_{hZZ}
&=& \epsilon_B \left ( \sqrt{2} G_F \right )^{1/2}
M_Z^2 Z_\mu Z^\mu h\\
{\cal L}_{hAZ}&=& - \epsilon_A \frac{g}{\cos\theta_W}
Z^\mu h \stackrel{\leftrightarrow}{\partial_\mu} A
\end{eqnarray}
The couplings $\epsilon_{A(B)}$ are model dependent.  For instance,
the SM Higgs sector has $\epsilon_A=0$ and $\epsilon_B=1$, while a
majoron model with one doublet and one singlet leads to $\epsilon_A=0$
and $\epsilon_B^2 \leq 1$.

The signatures of the Bjorken process and the associated production
depend upon the allowed decay modes of the Higgs bosons h and A. For
Higgs boson masses $m_h$ accessible at LEP II energies the main decay
modes for the CP-even state $h$ are $b \bar{b}$ and $JJ$. We treat the
branching fraction $B$ for $h \rightarrow JJ$ as a free parameter.
In most models $B$ is basically unconstrained and can vary from 0
to 1. Moreover, we also assume that, as it happens in the simplest
models, the branching
fraction for $A \rightarrow b \bar{b}$ is nearly one, and the
invisible $A$ decay modes $A\rightarrow hJ$, $A\rightarrow JJJ$ do not
exist (although CP-allowed).  Therefore our analysis
depends finally upon five parameters: $M_h$, $M_A$, $\epsilon_A$,
$\epsilon_B$, and $B$. This parameterisation is quite general and very
useful from the experimental point of view: limits on $M_h$, $M_A$,
$\epsilon_A$, $\epsilon_B$, and $B$ can be later translated into
bounds on the parameter space of many specific models.

The parameters defining our general parametrisation can be constrained
by the LEP I data. In fact, Refs.\ \cite{valle:1,valle:asso} analyse
some signals for invisible decaying Higgs bosons, and conclude that
LEP I excludes $M_h$ up to 60 GeV provided that $\epsilon_B > 0.4$.

The $\bar{b}b + \ptmis$ topology is our main subject of investigation
and we evaluate carefully signals and backgrounds, choosing the cuts
that enhance the signal over the backgrounds.  Our goal is to evaluate
the limits on $M_h$, $M_A$, $\epsilon_A$, $\epsilon_B$, and $B$ that
can be obtained at LEP II from this final state.  There are three
sources of signal events with the topology $\ptmis +$ 2 $b$-jets: one
due to the associated production and two due to the Bjorken mechanism.
\begin{eqnarray}
e^+  e^- &&\rightarrow ( Z  \rightarrow b   \bar{b} )~
+ ~(h \rightarrow  J  J)
\label{h:jj}
\\
e^+  e^- &&\rightarrow ( Z  \rightarrow \nu \bar{\nu} )~
+ ~(h \rightarrow  b \bar{b})
\label{h:sm}
\\
e^+  e^- &&\rightarrow ( A  \rightarrow b   \bar{b} )~
+ ~(h \rightarrow  J  J) \; .
\label{h:a}
\end{eqnarray}
The signature of this final state is the presence of two jets
containing $b$ quarks and missing momentum ($\ptmis$).  It is
interesting to notice that for light $M_h$ and $M_A$, the associated
production dominates over the Bjorken mechanism \cite{valle:asso}.

There are several sources of background for this topology:
\begin{eqnarray}
e^+ e^-  &&\rightarrow Z/\gamma~ Z/\gamma \rightarrow q \bar{q} ~ \nu \bar{\nu}
\label{zz:bbinv}
\\
e^+ e^-  &&\rightarrow (e^+e^-) \gamma\gamma \rightarrow [e^+e^-] q \bar{q}
\label{gg:bbinv}
\\
e^+ e^-  &&\rightarrow Z^*/\gamma^*~  \rightarrow q \bar{q} [n \gamma ]
\label{z:bbinv}
\\
e^+ e^-  &&\rightarrow W^+ W^- \rightarrow q \bar{q}^\prime ~ [\ell] \nu
\label{ww:bbinv}
\\
e^+ e^-  &&\rightarrow W [e] \nu \rightarrow q \bar{q}^\prime ~ [e] \nu
\label{ewn:bbinv}
\\
e^+ e^-  &&\rightarrow Z  \nu  \bar{\nu} \rightarrow q \bar{q} ~ \nu \bar{\nu}
\label{znn:bbinv}
\end{eqnarray}
where the particles in square brackets escape undetected and the jet
originating from the quark $q$ is identified (misidentified) as being
a $b$-jet.

At this point the simplest and most efficient way to improve the
signal-over-background ratio is to use that the Higgs bosons $A$ and
$h$ decays lead to jets containing $b$-quarks. So we require that the
events contain two $b$-tagged jets. Moreover, the background can be
further reduced requiring a large $\ptmis$. Having these facts in mind
we impose the following set of cuts, based on the ones used by the
DELPHI collaboration for the SM Higgs boson search~\cite{DELPHI}:

\begin{enumerate}
\item Charged multiplicity cut. We require that the event should
  contain more than 8 charged particles. With this cut we eliminate
  potential backgrounds from the production of $\tau^+ \tau^-$ pairs.
\item Missing momentum cuts. We require:
  \begin{itemize}
  \item The $z$ component of the missing momentum to be smaller than
    $0.15\times \sqrt{s}$.
  \item The absolute value of cosine of the polar angle of the missing
    momentum to be less than 0.9.
  \item The transversal component of missing momentum $\ptmis$ should be
    bigger than 25 GeV for $\sqrt{s}=175$ and $190$ GeV and 30 GeV for
    $\sqrt{s}=205$ GeV.
\end{itemize}
\item Acolinearity cut. The cosine of the angle between the axes of
  the two most energetic jets is required to be above -0.8. This is
  equivalent to the requirement that the angle between the axes is
  smaller than $145^{\circ}$.
\item Scaled acoplanarity cut. The scaled acoplanarity is computed as
  the complement of the angle in the perpendicular plane to the beam
  pipe between the total momenta in the two thrust hemispheres,
  multiplied by min $\left\{ \sin\theta_{jet~1}, \sin\theta_{jet~2},
  \right \}$ in order to remove instability at low polar jet
  angles~\cite{DELPHI}. Scaled acoplanarity is required to be greater
  than $7^{\circ}$.
\item Thrust/number of jets cut. We require the event thrust to be
  bigger than 0.8. For the intermediate visibly decaying Higgs boson
  masses in the range $45-80$ GeV this cut gives relatively small
  signal efficiency.  For this mass range instead of the thrust cut we
  demand that the two most energetic jets should carry more than 85\%
  of the visible energy.
\item Invariant mass cut. We assume that the visible mass should be in
  the range $M \pm 10$ GeV, where $M$ is the mass of the visibly
  decaying particle ($Z$, $h$, or $A$).
\item $b$-tagging cut. We adopt the efficiencies for the $b$-tagging
  directly from the DELPHI note~\cite{DELPHI}: 68\% efficiency
  for the signal and the appropriate values for the backgrounds
  extracted from Table 5 of ref.~\cite{DELPHI}.
\end{enumerate}

Depending on the $h$ and $A$ mass ranges, including or excluding the
invariant mass cut gives better or weaker limits on the $ZhA$ and
$ZZh$ couplings.  Therefore, for each mass combination four limits are
calculated (with or without invariant mass cut, with thrust cut or the
cut on the minimal two-jet energy) and the best limit is kept.

We denote the number of signal events for the three production processes
(\ref{h:jj} -- \ref{h:a}), after imposing all cuts, $N_{JJ}$, $N_{SM}$,
and $N_A$ respectively, assuming that $\epsilon_A = \epsilon_B = 1$. Then
the expected number of signal events when we take into account couplings
and branching ratios is
\begin{equation}
N_{exp} = \epsilon_B^2 \left [ B N_{JJ} + (1-B) N_{SM} \right ]
+ \epsilon_A^2 B N_A \; .
\end{equation}
In general, this topology is dominated by the associated production,
provided it is not suppressed by small couplings $\epsilon_A$ or phase
space. The most important background after the cuts is
(\ref{zz:bbinv}). The total numbers of background events summed
over all relevant channels are 2.3, 2.8
and 5.9 for $\sqrt{s}=175$ , $190$ and $205$ GeV respectively.

\begin{figure}[htbp]
\vspace*{-15mm}
\begin{center}
\begin{tabular}{p{0.48\linewidth}p{0.48\linewidth}}
\begin{center}
\mbox{\epsfig{%
file=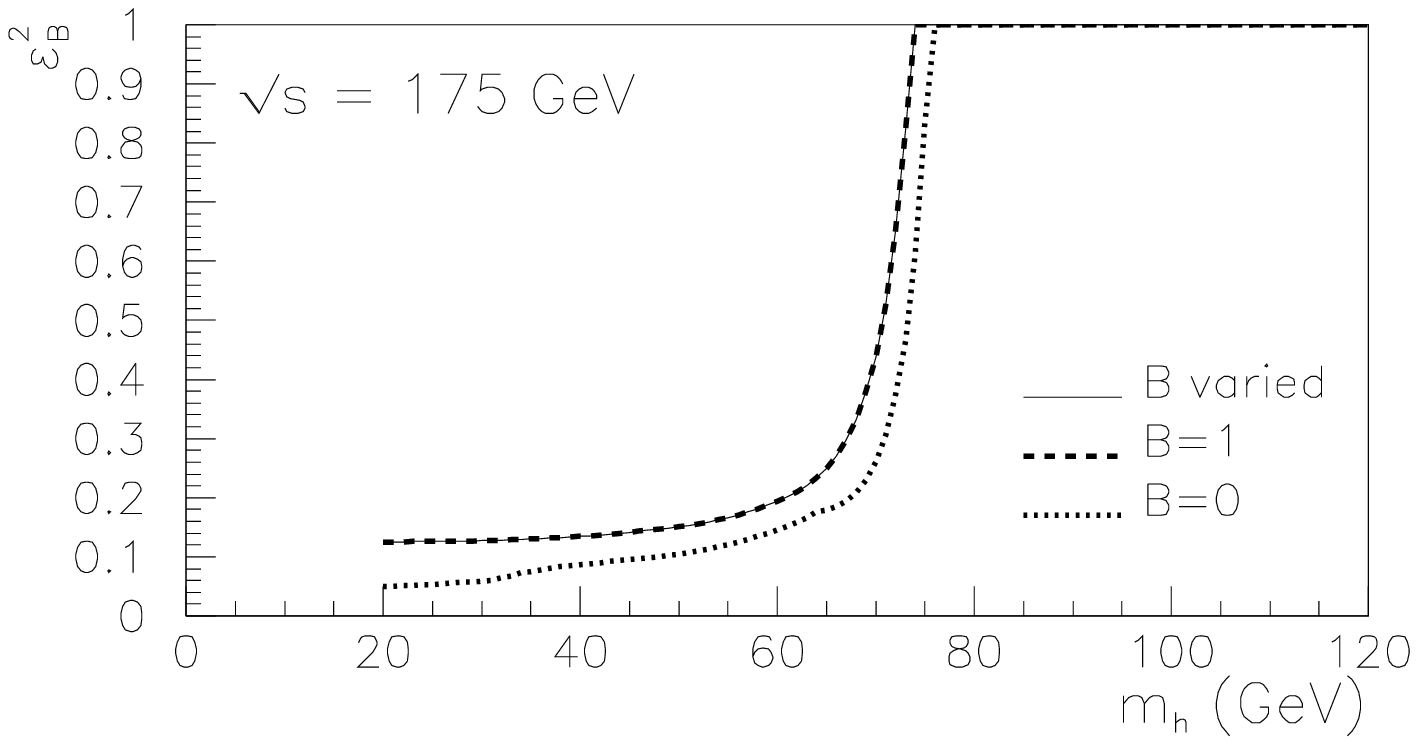,width=\linewidth,bbllx=0pt,bblly=10pt,bburx=410pt,bbury=235pt}}
\end{center}&
\begin{center}
\mbox{\epsfig{%
file=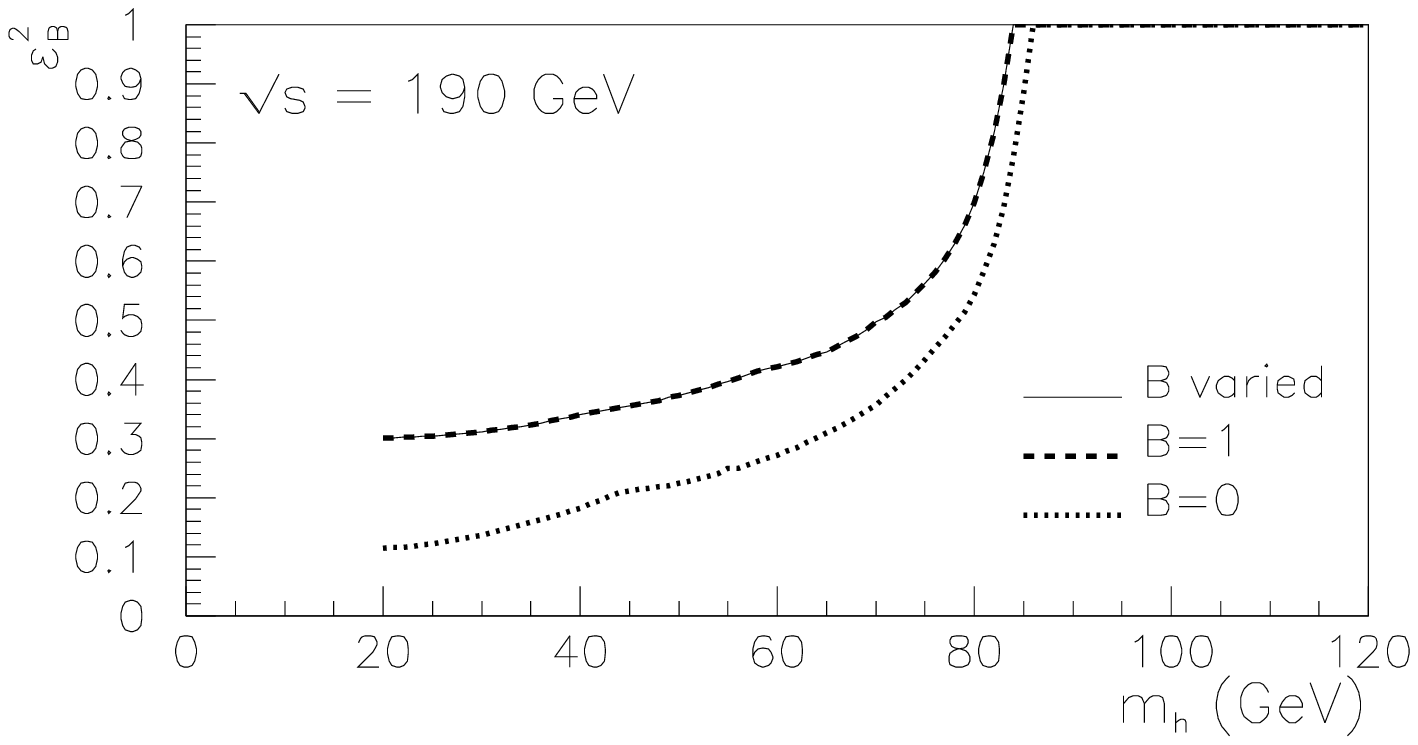,width=\linewidth,bbllx=0pt,bblly=10pt,bburx=410pt,bbury=235pt}}
\end{center}\\
\end{tabular}
\vspace*{-13mm}
\caption{\protect\small
\baselineskip 12pt
Limits on $\epsilon_B^2$ as a function of $M_h$ for $\surd{s}=175,190$
GeV and for different values of $B=Br(h\rightarrow JJ)$}
\label{fig:zh}
\end{center}
\vspace*{-8mm}
\end{figure}

\begin{figure}[htbp]
\vspace*{-10mm}
\begin{center}
\begin{tabular}{p{0.48\linewidth}p{0.48\linewidth}}
\begin{center}
\mbox{\epsfig{%
file=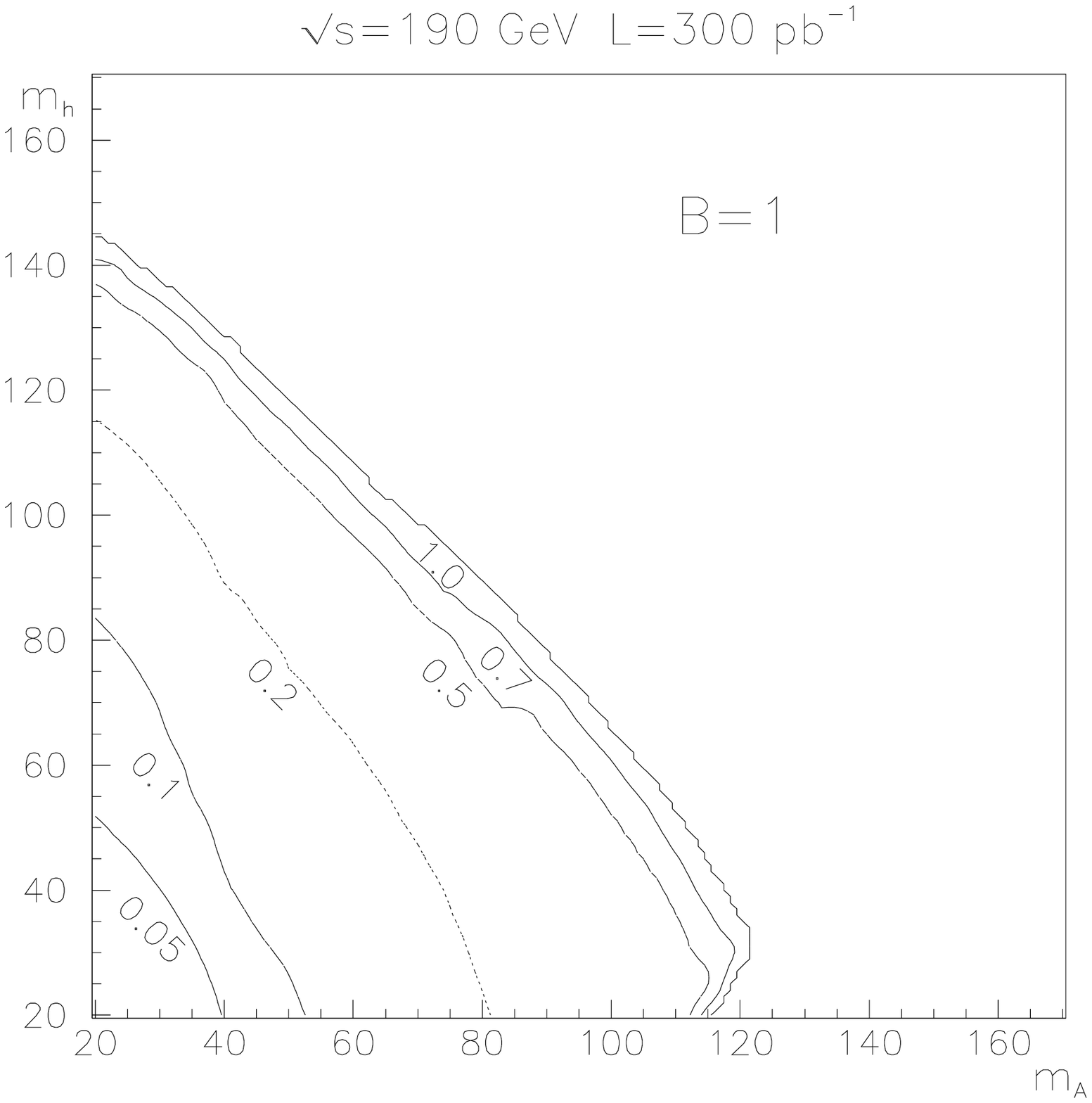,width=\linewidth,bbllx=10pt,bblly=10pt,bburx=530pt,bbury=530pt}
}\end{center}&
\begin{center}
\mbox{\epsfig{%
file=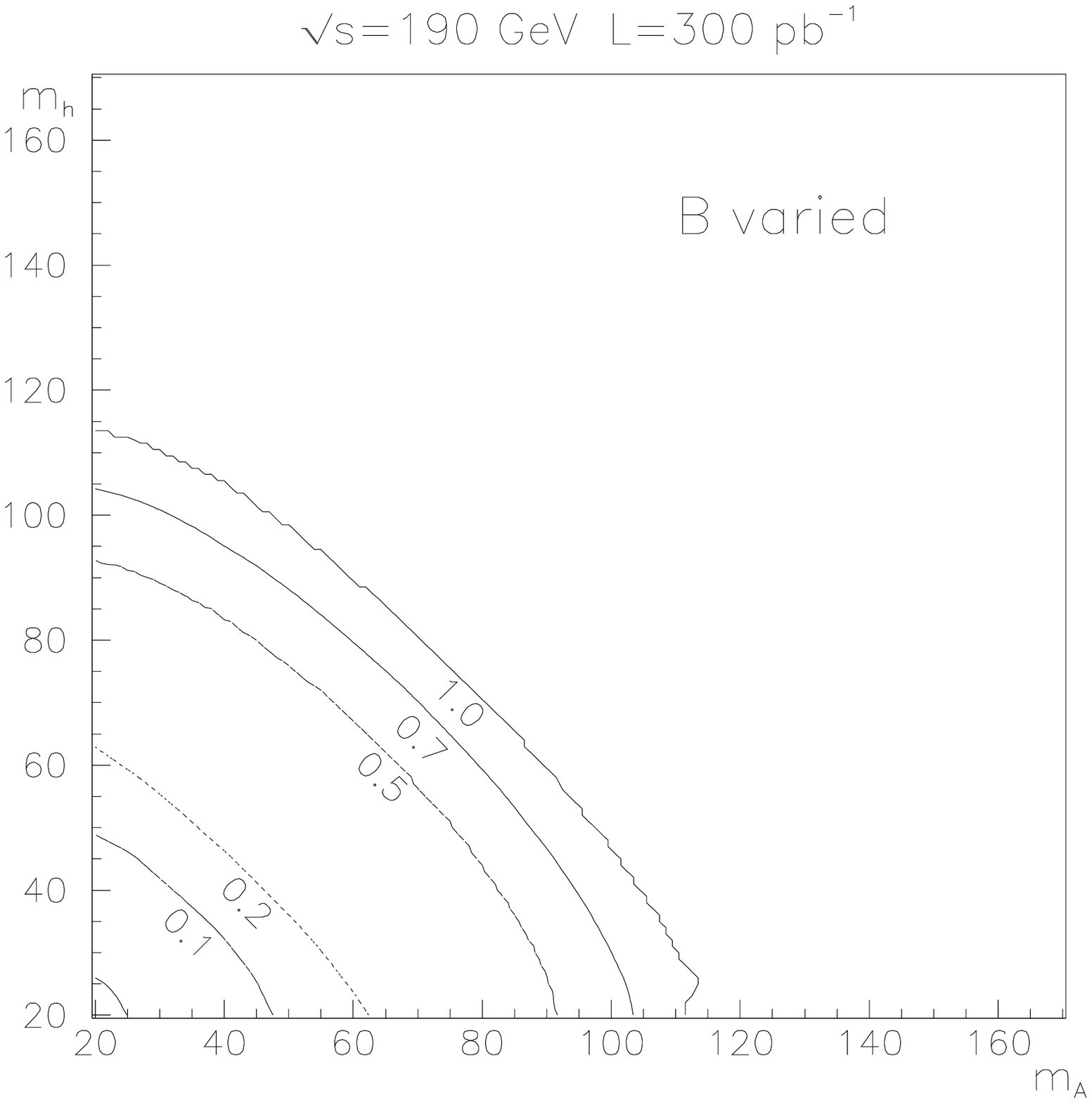,width=\linewidth,bbllx=10pt,bblly=10pt,bburx=530pt,bbury=530pt}}
\end{center}\\
\end{tabular}
\vspace*{-10mm}
\caption{\protect\small
\baselineskip 12pt
Limits on $\epsilon_A^2$ as a function of $M_h, M_A$ for
$\surd{s} = 190$ GeV. The left plot shows the limits obtained
for $B=Br(h\rightarrow JJ)=1$, in the right plot $B$ is varied from 0 to 1.}
\label{fig:ah}
\end{center}
\vspace*{-5mm}
\end{figure}

In order to obtain the limits shown in
Figs.~\ref{fig:zh}-\ref{fig:ah}, we assumed that only the background
events are observed, and we evaluated the 95 \% CL region of the
parameter space that can be excluded with this result.  By taking
the weakest bound, as we vary $B$, we obtained the absolute bounds on
$\epsilon_A$ and $\epsilon_B$ independent of the $h$ decay
mode. The limits on $\epsilon_A$ obtained by searches for the
$b\protect\bar{b} ~+~ \protect\ptmis$ final states are stronger
than those given by the $b\protect\bar{b}b\protect\bar{b}$
topology.  The bounds on $\epsilon_B$  apply directly also for
the simplest model of invisibly decaying Higgs bosons, where just one singlet
is added to the SM. A more complete presentation of these results will be
given in ref.~\cite{inprep}.

\begin{center}
{\bf ACKNOWLEDGEMENTS}
\end{center}

This work was supported by the University of Wisconsin Research
Committee with funds granted by the Wisconsin Alumni Research
Foundation, by the U.S.\ Department of Energy under Grant
No.~DE-FG02-95ER40896, by DGICYT under Grant No.~PB92-0084, by Conselho
Nacional de Desenvolvimento Cient\'{\i}fico e Tecnol\'ogico
(CNPq/Brazil), by Funda\c{c}\~ao de Amparo \`a Pesquisa do Estado de
S\~ao Paulo (FAPESP/Brazil), and by a DGICYT postdoctoral fellowship
and the Alexander von Humboldt Stiftung. We thank S. Katsanevas for
useful discussions and for bringing the paper of ref.~\cite{DELPHI}
to our attention.

\vskip 1mm

\newpage
$^1$ Permanent address: Instituto de F\'{\i}sica, Universidade de S\~ao Paulo,
C.P.\ 66318, CEP 05389-970 S\~ao Paulo, Brazil

\end{document}